\documentclass[preprint,showpacs,preprintnumbers,amsmath,amssymb]{revtex4}

\usepackage{amssymb}
\usepackage{amsmath}
\usepackage{graphicx}

\begin{document}
\title{Topological view on  magnetic adatoms in graphene}

\author{Zhen-Gang Zhu, and Jamal Berakdar}
\affiliation{Institut f\"ur Physik, Martin-Luther-Universit\"at Halle-Wittenberg, 06099 Halle, Germany}

\begin{abstract}
We study theoretically the physical properties of a magnetic impurity in graphene. The theory is
based on the Anderson model with a very strong Coulomb interaction on the impurity.
We start from the Slave-Boson method and introduce a topological picture consisting of a
degree of a map and a winding number (WN) to analyze the phase shift and the occupation on the impurity. The occupation  is linked to  the WN. For  a generic normal metal we find a fractional WN. In contrast, the winding is accelerated by the relativistic dispersion of graphene at half-filling in which
case  an
integer occupation is realized.
We show that the  renormalization  that shifts the impurity level is insufficient  to invert the sign of the energy level. Consequently, the state at half-filling is stable unless a  gate voltage is tuned such that the Fermi level touches the edge of the broadened impurity level. Only in this case
 the zero field susceptibility is finite and shows a pronounced peak structure when scanning the gate voltage.
\end{abstract}


\maketitle

\section{Introduction}
 Graphene has attracted many theoretical and experimental researches due to its unique properties
  which are also of relevance for technological  applications  \cite{graphene,neto}. The hallmark
  of graphene, a monolayer of carbon atoms, is its electronic structure
with the valence and the conduction bands touching at two inequivalent  points $K_{\pm}$ at the corners of the first Brillouin zone (FBZ). The low energy dispersion around  $K_{\pm}$  is relativistic (linear in momentum), with a massless Dirac fermion behavior \cite{neto}.
Of a particular interest is the issue  of how
  the nature of graphene is manifested  in the behavior of  magnetic adatoms \cite{meyer,cornaglia,zhuang,kotov,uchoa,ding,uchoa0906,hentschel,dora,sengupta,jacob,dellanna,zhu10,zhu09,zhu10_2}, a topic at the heart of
  many-body physics,
   and especially  the Kondo effect \cite{kotov,uchoa,ding,uchoa0906,hentschel,dora,sengupta,jacob,dellanna,zhu10}.  The Kondo model with a linear dispersion \cite{withoff} and the Anderson model in $d$-wave superconductors \cite{zhang} have been  investigated already.
 As detailed below however, for the particular case of magnetic adatoms on graphene
 some additional features emerge.\\  The problem of adatoms on graphene was treated
     within the Hatree-Fock approximation \cite{uchoa,ding,uchoa0906}. This  is only valid at temperatures
$T>T_K$, where $T_K$ is the Kondo temperature. The anisotropic \emph{single} channel
Kondo model  \cite{hentschel} and the Anderson model \cite{dora} for  infinite Coulomb correlation ($U$)
were also considered. A Fermi liquid behaviour \cite{dora} were concluded.
In contrast, Ref. \cite{sengupta} arrives at  a two-channel Kondo in graphene
due to the \emph{valley degeneracy} of the Dirac electrons
leading to an over-screening and thus to a
non-Fermi-liquid-like ground state. In Ref. \cite{zhu10}, we conducted a detailed
symmetry group analysis  to clarify the appropriate physical model  and highlighted
 the  various relevant symmetries that are realized
 depending on whether (A) the adatom is  above one carbon atom or  (B) this atom is in the center of the honeycomb.
 The contributions from the two Dirac cones are mixed. For the case B we found
  generally a multi-channel, multi-flavor Kondo model. While for A  we inferred a one-channel, two-flavor behavior. To identify the correct starting Hamiltonian a   symmetry analysis is imperative.
  For example, the detailed symmetry analysis in \cite{zhu10} for the A and B cases yields that the realized symmetry groups are $\bar{C}_{3v}$ and $C_{6v}$ point groups. To be consistent with these symmetries, the eigenstates for pure graphene at Dirac points should be recombined as to reflect the modifications
    imparted by the impurity. As a consequence, a single half spin with zero orbital angular momentum is \emph{decoupled} from graphene in case B (in contrast Ref. \onlinecite{uchoa2011}).

In this work, we focus on the  situation A and consider a two-flavor Anderson model with a relativistic dispersion relation. Therefore,  the effect of the gate voltage   can be studied in a wide   spectrum  since the charge fluctuations are already taken into account \cite{footnote}.
We introduce a geometrical picture in form of a  winding number (WN) and a degree of a map, to analyze the occupation and the phase shift. It is shown  that the occupation takes on the values 0 or 1 when the bare level is respectively  above or below the Fermi energy at half-filling. In the presence of a gate voltage, our analytical and the numerical calculations show that the state remains stable only when the edge of the broadened impurity level touches the Fermi energy where a nonzero susceptibility occurs. These findings are traced back mainly to  the relativistic dispersion in graphene. In section II, an appropriate formulation  is worked out,  followed by the topological interpretation in the section III. In  section IV, numerical illustrations are displayed for the half-filling case and  beyond. In  section V, the   renormalization of the impurity level and the occupation dependence on the varying gate voltage are calculated.  We conclude with a summary of this study.

\section{Framework}
%Keeping the nearest neighbor hopping, we get the
We start from the Anderson Hamiltonian \cite{zhu10}
$$H=H_{\text{g}}+H_{\text{hyb}}+H_{\text{im}}$$
where the terms $H_{\text{g}}$, $H_{\text{hyb}}$, and $H_{\text{im}}$ describe respectively  graphene,
the hybridization, and the impurity. %In a single particle picture we write
The graphene Hamiltonian reads \cite{zhu10} $$H_{\text{g}}=\sum_{s\sigma}\int_{-k_{c}}^{k_{c}}dk\varepsilon_{k}c_{sk\sigma}^{\dagger}c_{sk\sigma},$$
where $\varepsilon_{k}=\hbar v_{F}k$, and $v_{F}$ is the Fermi velocity, $s$ and $\sigma$ are
 valley and spin indices. $k_{c}$ is the cut-off momentum  that sets  the linear
 dispersion region. $c_{sk\sigma}$ is an annihilation operator of the one electron state $|sk\sigma\rangle$.
The Hamiltonian of the impurity is treated  in the $U\rightarrow\infty$ limit,  in a standard way \cite{hewson,coleman}.
We introduce a bosonic field $b$ to guarantee  the ($Q$) charge conservation
%\begin{equation}
$$Q=b^{\dagger}b+n_{f}=1,$$
%\label{q}
%\end{equation}
%
where $$n_{f}=\sum_{\sigma}f_{\sigma}^{\dagger}f_{\sigma},$$  and $f_{\sigma}$ is the annihilation
operator of an electronic state on the impurity with spin $\sigma$.
Within the slave-boson (SB) model  $H$ reads then
%\begin{equation}
$$H=\tilde\varepsilon_{0}n_{f}+H_{\text{g}}+H_{\text{hyb}}+\lambda(b^{\dagger}b-1).$$
%\label{htilde}
%\end{equation}
Here the normalized impurity energy level is $$\tilde\varepsilon_{0}=\varepsilon_{0}+\lambda.$$
The hybridization Hamiltonian is formulated as
\begin{equation}
H_{\text{hyb}}=v_{0}\sqrt{\pi\Omega_{0}}\sum_{s\sigma}
\left(\int\frac{\sqrt{|k|}dk}{2\pi}c_{sk\sigma}^{\dagger}b^{\dagger}f_{\sigma}+h.c.\right).
\label{hhyb1}
\end{equation}
 $v_{0}$ is the hybridization strength, and $\Omega_{0}$ is the area of a unit cell.
As usual the bosonic field is assumed to be condensed at the ground state and is described by
 a renormalization number as $$\langle b^{\dagger}\rangle=\langle b\rangle=\zeta.$$
   $\lambda$ and $\zeta$ are determined by minimizing the free energy which leads to the equations $$\zeta^{2}=1-\langle n_{f}\rangle$$ and
\begin{equation}
\lambda=-\zeta^{-1}v_{0}\sqrt{\pi\Omega_{0}}\sum_{s\sigma}\int_{-k_{c}}^{k_{c}}dk\frac{\sqrt{|k|}}{2\pi}\langle f_{\sigma}^{\dagger}c_{sk\sigma}\rangle.
\label{conditions}
\end{equation}
The Green's function associated with the impurity is
\begin{eqnarray}
G_{f\sigma}=z^{-1},  \mbox{ where} \quad
z=\omega^{+}-\tilde\varepsilon_{0}-\zeta^{2}\Sigma_{0}(\omega^{+}),\: \omega^{+}=\omega+i\delta,\; \delta\rightarrow0^{+} .
\label{gf}
\end{eqnarray}
 The selfenergy is defined as
$$\Sigma_{0}(\omega^{+})=(v_{0}^{2}\Omega_{0}/4\pi)\sum_{s}\int dk|k|(\omega^{+}-\varepsilon_{k})^{-1},$$
and can be given analytically as
\begin{equation}
\Sigma_{0}=-\frac{N_{s}v_{0}^{2}}{2}
\left[\rho\ln\frac{|D^{2}-\omega^{2}|}{\omega^{2}}+i\pi|\rho|\theta(D-|\omega|)\right]. \label{selfenergy1}
\end{equation}
 The density of state of  graphene around the Dirac points reads \cite{neto} $$\rho(\omega)=\frac{\Omega_{0}}{2\pi}\frac{\omega}{(\hbar v_{F})^{2}}.$$
  %We stress that the selfenergy is valid when $\omega\in (-D,D)$.
Defining the local density of states (LDOS) of the impurity as $$\mathcal{N}_{f\sigma}=-\frac{\Im}{2\pi}G_{f\sigma},$$ we find
\begin{equation}
\langle n_{f}\rangle=2\int_{-D}^{\varepsilon_{F}}f(\omega)\mathcal{N}_{f\sigma}d\omega,
\label{occup}
\end{equation}
where $f(\omega)$ is the Fermi function. Following \cite{hewson}, we derive
\cite{footnote2}
\begin{equation}
\langle n_{f}\rangle=-\frac{1}{\pi}\Im\int_{-D}^{\varepsilon_{F}}d\ln z.
\label{occup1}
\end{equation}
Here  we introduced $$z=r\exp{i[\pi/2-\Theta(\omega,\lambda,\zeta^{2})]},$$ and
\begin{equation}
\Theta(\omega,\lambda,\zeta^{2})=\tan^{-1}
\left[\frac{\omega-\tilde{\varepsilon}_{0}-\zeta^{2}\Re\Sigma_{0}}{-\zeta^{2}\Im\Sigma_{0}}\right]
\label{phase}
\end{equation}
is the phase of $z$ in $(-\pi/2,\pi/2)$.
Therefore, the occupation number reads
\begin{equation}
\langle n_{f}\rangle=\frac{1}{2\pi}\int_{-\infty}^{\varepsilon_{F}}d\left(2\Theta(\omega,\lambda,\zeta^{2})\right)=\text{deg(the map)},
\label{occup2}
\end{equation}
which is  the Friedel sum rule for an impurity on graphene.

\section{Topological interpretations}
%
%Please note that the lower limit in the integral is extended from $-D$ to $-\infty$ since $\omega=-D$ is a %singular point. This extension leads to a dramatic decaying of the selfenergy (see Eq. (\ref{selfenergy1})) in %the extended region.
Now we introduce a new picture that allows a
topological interpretation by extending the concept of the degree of a map and a WN of a closed curve to an open curve.
In Eq. (\ref{occup2}), $2\Theta(\omega,\lambda,\zeta^{2})$ defines a map: $O \mapsto P$, where $O$ and $P$ are both 1D manifolds. $O$ stands for the 1D energy region from $-\infty$ to $D$; and the manifold $P$ is ${\cal S}^{1}$.
 %If we use $D/\pi$ as the unit of the energy, we then convert the $O$ into another $S^{1}$, from $-\pi$ to %$\pi$. By defining $\phi=\pi\omega/D$, $\phi_{F}=\pi\varepsilon_{F}/D$, %$\tilde\phi_{0}=\pi\tilde{\varepsilon}_{0}/D$, $\phi_{R\Sigma}=\pi\Re\Sigma_{0}/D$, and  %$\phi_{I\Sigma}=\pi\Im\Sigma_{0}/D$, we obtain
%
%
%\begin{equation}
%\langle n_{f}\rangle=\frac{1}{2\pi}\int_{-\pi}^{\phi_{F}}d\left(2\Theta(\phi,\lambda,\zeta^{2})\right)=\text{deg(the map)},
%\label{occup3}
%\end{equation}
%
%
%and
%
%\begin{equation}
%\Theta(\phi,\lambda,\zeta^{2})=\tan^{-1}\left[\frac{\phi-\tilde\phi_{0}-\zeta^{2}\phi_{R\Sigma}}{-\zeta^{2}\phi_{I\Sigma}}\right].
%\label{phase1}
%\end{equation}
%
The integral in (\ref{occup2}) can be viewed as a  winding process by varying the source point $\bar{p}$ (stands for $\varepsilon_{F}$). Simultaneously, the image point $p$ scans in the manifold P shown in Fig. (1a). If $p$ has a cyclic winding, usually an integer for the number of times that the manifold $O$ covers the manifold $P$ is produced and  called a WN (similar to the case of a continuous map \cite{dubrovin}). Note, our degree of the map $2\Theta$  has the same topological meaning for non-integer WN, it indicates then that the winding process is not complete. We note further,   $\Theta$ is the connection in the image manifold $P$, and $\partial\Theta/\partial\omega$ is the curvature of this manifold.
\begin{figure}
\includegraphics[width=0.25\columnwidth,angle=270]{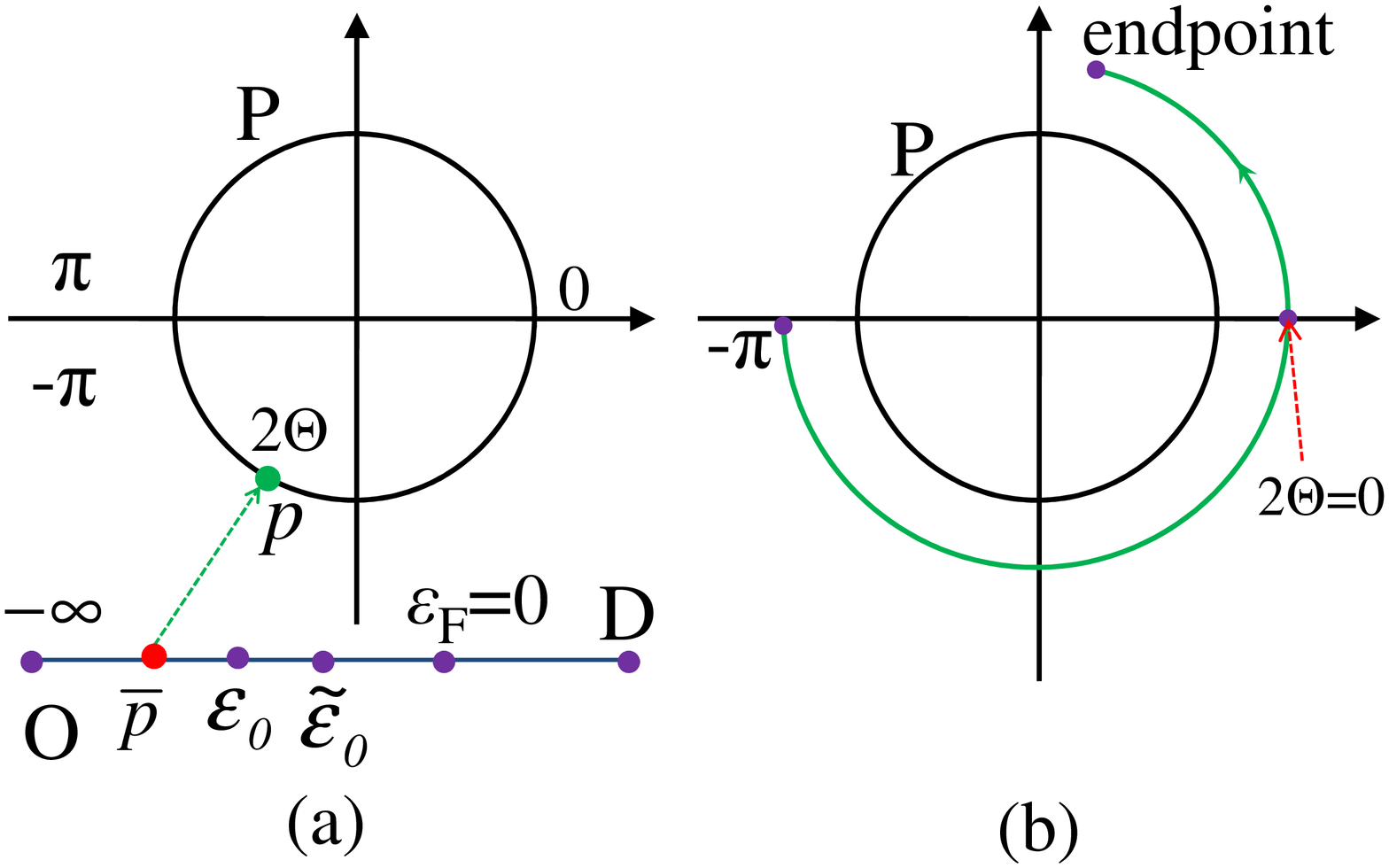}
\caption{(color online) (a) Schematics of  the map from 1D manifold O to P. (b) The winding process for a normal metal. \label{f1}}
\end{figure}

For a comparison, let us recall the same map for a normal metal with a constant DOS ($\omega\in(-\infty,\varepsilon_{F}]$), i.e. $$\Theta(\omega)=\tan^{-1}\left((\omega-\varepsilon_{0})/\Delta_{0}\right).$$ %where $\Delta_{0}=\pi v_{0}^{2}\rho$ is the imaginary part of the selfenergy for a normal metal.
When $\bar{p}$ reaches  $\varepsilon_{0}$, the image point $p$ attains $\Theta=0$, meaning a half winding of the manifold $P$. When $\bar{p}$ moves over $\varepsilon_{0}$ and  approaches  the Fermi level $\varepsilon_{F}$, the image point $p$ stops somewhere in the upper branch of the manifold $P$ if $2\Theta\neq\pi$ (see the endpoint in Fig. (1b)). The winding is not completed so that the occupation number is not an integer but a fractional number. In the case that a mean-field SB method is applied and the real part of the selfenergy is ignored, the map reads
 $$\Theta(\omega)=\tan^{-1}\left(\frac{\omega-\tilde{\varepsilon}_{0}}{\zeta^{2}\Delta_{0}}\right).$$ If $\zeta$ is not zero, $\varepsilon_{0}$ is renormalized to become $\tilde{\varepsilon}_{0}$ which is mapped onto  $\Theta=0$. The $\lambda$ in $\tilde{\varepsilon}_{0}$ gives rise to a shift of the impurity level, and $\zeta^{2}$ changes the winding velocity of the image point.

For the graphene case, we define a function
\begin{equation}
\mathcal{F}(\omega,\lambda,\zeta^{2})=\omega-\zeta^{2}\Re\Sigma_{0}(\omega^{+}),
\label{equation}
\end{equation}
which is shown in Fig. (2a). The starting point of the winding is fixed at $2\Theta=-\pi$ since $\mathcal{F}\rightarrow -\infty$ as $\omega\rightarrow -\infty$. In Fig. (2a), the solid horizontal line indicates the position of $\tilde{\varepsilon}_{0}$ so that the crossing points with $\mathcal{F}$ indicate the zeros ( $2\Theta=0$ ) of the image points $p$ and the corresponding pre-image points (source points), i.e. $\omega^{*}$s of $\bar{p}$. Counting the number of the pre-image points from the inverse map, generally, there are two points, i.e. $\omega^{*}_{1},\omega_{2}^{*}$, residing closely to $\omega=-D$ in two separated  regions, I and II, with opposite countings of the degree of the map. The index of $\omega^{*}_{1}$ ($\omega_{2}^{*}$) is $+1$ ($-1$). The behavior of $\omega=0$ (Dirac point at half-filling) is particularly  important for our analysis. As the imaginary part of the selfenergy diminishes, the end of the image point for $\omega=0$ is  determined by the relative positions of the function $\mathcal{F}(\omega=0,\lambda,\zeta^{2})$ and $\tilde{\varepsilon}_{0}$. When the former is larger (less) than the latter, the endpoint is $\pi$ ($-\pi$). The linear dispersion around the Dirac points accelerates the phase shift so that the winding process  speeds up to the boundary of the manifold $P$.

\begin{figure}%[tph]
\includegraphics[width=0.35\columnwidth, angle=270]{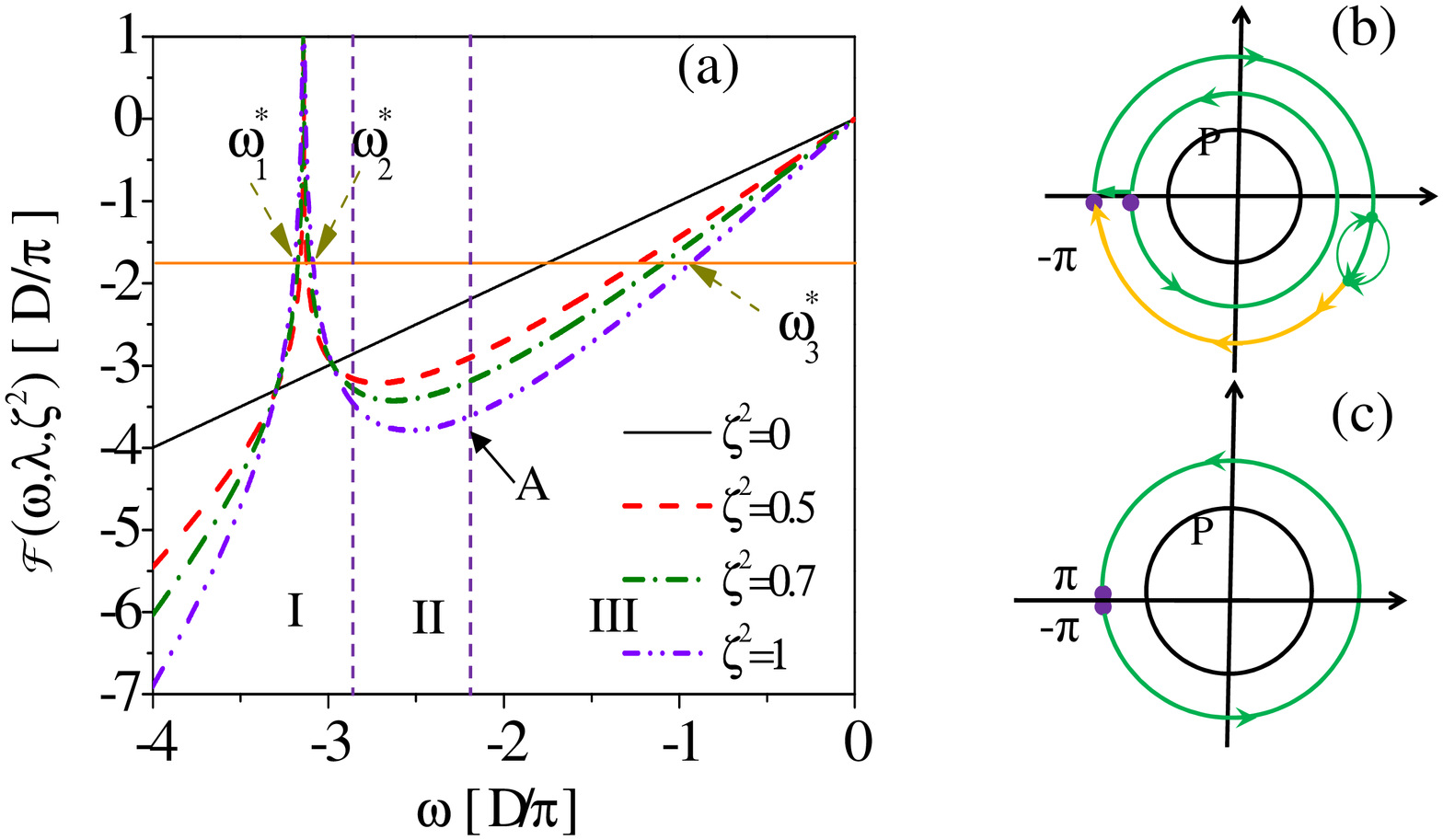}
\caption{(color online) (a) Numerical graphs for $\mathcal{F}(\omega,\lambda,\zeta^{2})$ and (b) a schematic of the winding process for $\zeta^{2}=1$, $\tilde{\varepsilon}_{0}>0$.  In (c)  $\zeta^{2}=0$, $\tilde{\varepsilon}_{0}<0$. The energy unit is $D/\pi$. In (a) $\eta=\frac{\Omega_{0}}{2\pi}\frac{v_{0}^{2}}{(\hbar v_{F})^{2}}$ is assumed to be 0.4.} \label{f2}
\end{figure}

A schematic diagram of such a winding for $\zeta^{2}=1$ and $\tilde{\varepsilon}_{0}>0$ is shown in Fig. (2b). While the first  cycle is finished in region (I), in region (II)  $p$ comes back from $+\pi$  clockwise  to "A" point, approximately crossing  the zero of $2\Theta$ (at $\omega_{2}^{*}$). In region (III),  $p$ starts approximately from "A" winding  anticlockwise. However, it does not reach the zero of $2\Theta$ and turns back to $-\pi$ to finish the winding. Hence, a zero winding is concluded for such a case, which is consistent with  the condition of minimizing the free energy.  Fig. (2c) shows the winding for zero $\zeta^{2}$ and $\tilde{\varepsilon}_{0}<0$ where the WN is 1. We should note that zero $\zeta$ does not mean that
there is no effect from graphene to the impurity state. The charge fluctuation  still renormalizes the impurity energy level via the parameter $\lambda$ in this case. %In Fig. (2d), one-half winding is shown as the endpoint is $\Theta=0$. This %situation can only be achieved when $\tilde{\varepsilon}_{0}\rightarrow0$ and is less than $\omega$ at least by one order. If $\tilde{\varepsilon}_{0}$ violates this %restriction and has a slight shift from zero, the image point will rotate to the endpoints ($\pm\pi$) immediately and the winding will be completed instantaneously. It %seems to conclude that this point is not stable since charge fluctuation may lead to the fluctuation of the impurity level via parameter $\lambda$.

%The winding number is only determined by the ending point $\phi_{F}$, so does $\zeta^{2}$. For the state (I), $\mathcal{F}(\phi_{F})<\tilde{\phi}_{0}$, zero occupation %($\zeta^{2}=1$) is achieved shown by the black-solid line in Fig. (2a). In state (II), $\mathcal{F}(\phi_{F})=\tilde{\phi}_{0}$, occupation is half ($\zeta^{2}=1/2$) %shown by the blue-dotted line. In state (III), $\mathcal{F}(\phi_{F})>\tilde{\phi}_{0}$, occupation is one ($\zeta^{2}=0$) shown by the red-dashed line.

It is readily  shown that the occupation  takes on only  the values $0$ or $1$ when $\tilde{\varepsilon}_{0}>0$ or $<0$.
 %($\lambda$ is shown not to be sufficient large in the following).
 For instance, starting  initially from $\varepsilon_{0}<0$, $\lambda=0$ and $\zeta^{2}=0$ we obtain the occupation 1. Thus, $\zeta^{2}=0$ is stable against the renormalization process. Starting with $\varepsilon_{0}<0$, $\lambda=0$ and $\zeta^{2}\neq0$, we find the occupation  1 after one step. This leads to a new $\zeta^{2}=0$. Therefore, we may conclude that the renormalization in graphene is quite different to that in normal metals. When the bare level is below the Fermi level
  %the impurity is immediately localized to against the environment of graphene and
  the renormalization effect is small.

\section{Numerical illustrations and the case beyond half-filling}
Let us consider the effect of a finite gate voltage $v_{\text{g}}$ when the system is away from half-filling. To calculate the susceptibility, we introduce a homogenous, static external magnetic field
% within Zeeman term and take a vanishing field
and let it tend to zero at the end of calculations so that a zero-field susceptibility is obtained.
The occupation number and the magnetization of the impurity  read
$$\langle n_{f} \rangle= \frac{1}{2\pi}\sum_{\sigma}\int_{-\infty}^{v_{\text{g}}}d\Theta_{\sigma}(\omega,\lambda,\zeta^{2}),$$ and
$$M_{f}=\frac{\mu_{B}}{2\pi}\sum_{\sigma}\int_{-\infty}^{v_{\text{g}}}\sigma d\Theta_{\sigma}(\omega,\lambda,\zeta^{2}),$$
where the phase $\Theta_{\sigma}$ is now spin-dependent containing  the spin-dependent $\Sigma_{0\sigma}$, and $$\tilde{\varepsilon}_{0\sigma}=\tilde{\varepsilon}_{0}-\sigma h,\, h=\frac{1}{2}g\mu_{B}B,$$ where $g$ is the Land\'{e} factor, $\mu_{B}$ is the Bohr magneton. To investigate the stability of $\zeta^{2}=0$ at non-half-filling, it is crucial to determine the position of $\lambda$.
\begin{figure}
\includegraphics[width=9 cm, height=7 cm]{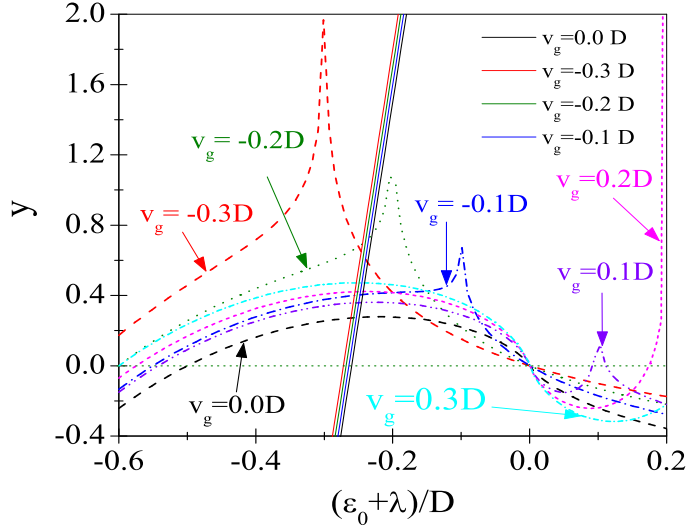}
\caption{(color online) Determination of  $\lambda$. The solid lines correspond  to $y_{1}=(\tilde{\varepsilon}_{0}-\varepsilon_{0})/2\eta D-(1+v_{\text{g}}/D)$. The dashed curves are calculated from  $\ln$ terms in Eq. (\ref{lambdavgh}). The dotted lines are grid lines. The energy unit is $D$. The other parameters are $\varepsilon_{0}=-0.3$ D, $\eta=0.02$.  \label{fig3}}
\end{figure}
We find, for $\zeta^{2}=0$,
\begin{equation}
\lambda=\left\{
\begin{array}{c}
\eta\left[2(D+v_{\text{g}})+\sum_{\sigma}\tilde{\varepsilon}_{0}
\ln\frac{|v_{\text{g}}-\tilde{\varepsilon}_{0\sigma}|}{\tilde{\varepsilon}_{0\sigma}+D}\right],  \\
\eta\left[2(D-v_{\text{g}})+\sum_{\sigma}\tilde{\varepsilon}_{0}\ln\frac{\tilde{\varepsilon}^{2}_{0}}{|\tilde{\varepsilon}_{0\sigma}+D||v_{\text{g}}
-\tilde{\varepsilon}_{0\sigma}|}\right],
\end{array}
\right.
\label{lambdavgh}
\end{equation}
for $v_{\text{g}}+\sigma h\leq 0$,  and $>0$ respectively. We solve $\lambda$  graphically, as shown in Fig. 3. The crossing points of the solid lines and the dashed lines deliver the solutions of $\lambda$ for a  given gate voltage. The logarithmic terms show an interesting behaviour, the peaks being present at the positions of the gate voltages (i.e. the positions of the Fermi levels). This graph differs substantially from that discussed by  Lacroix \cite{lacroix} for a normal metal. A good approximate solution of $\lambda$ is inferred by replacing the renormalized impurity level in Eq. (\ref{lambdavgh}) with $\varepsilon_{0}$. As known, the renormalization to the energy level by the hybridization is unlikely to change the sign of the level.
The occupation can not be changed by  $\lambda$ alone.
 As a consequence, $\zeta^{2}=0$ is stable as long as $\tilde{\varepsilon}_{0\sigma}<v_{\text{g}}$. For
  a finite  $\zeta^{2}$ we resort to numerical calculations. We derive the susceptibility (higher orders in $v_{\text{g}}$ are ignored)
\begin{equation}
\chi=\frac{\eta\zeta^{2}\mu^{2}_{\text{B}}}{2}\sum_{\sigma}\frac{\text{sgn}(v_{\text{g}\sigma})\left(\tilde{\varepsilon}_{0}-\eta\zeta^{2}v_{\text{g}\sigma}\right)}
{(\eta\zeta^{2}\pi)^{2}v^{2}_{\text{g}\sigma}+(\tilde{\varepsilon}_{0}-v_{\text{g}\sigma}+\zeta^{2}\Re\Sigma_{\sigma})^{2}},
\label{chi}
\end{equation}
where $v_{\text{g}\sigma}=v_{\text{g}}+\sigma h$. Interestingly  $\chi$ changes its sign in accordance with the sign-change of $v_{\text{g}\sigma}$ which reflects the particle-hole nature of the graphene.
\begin{figure}
\includegraphics[width=0.45\columnwidth, angle=270]{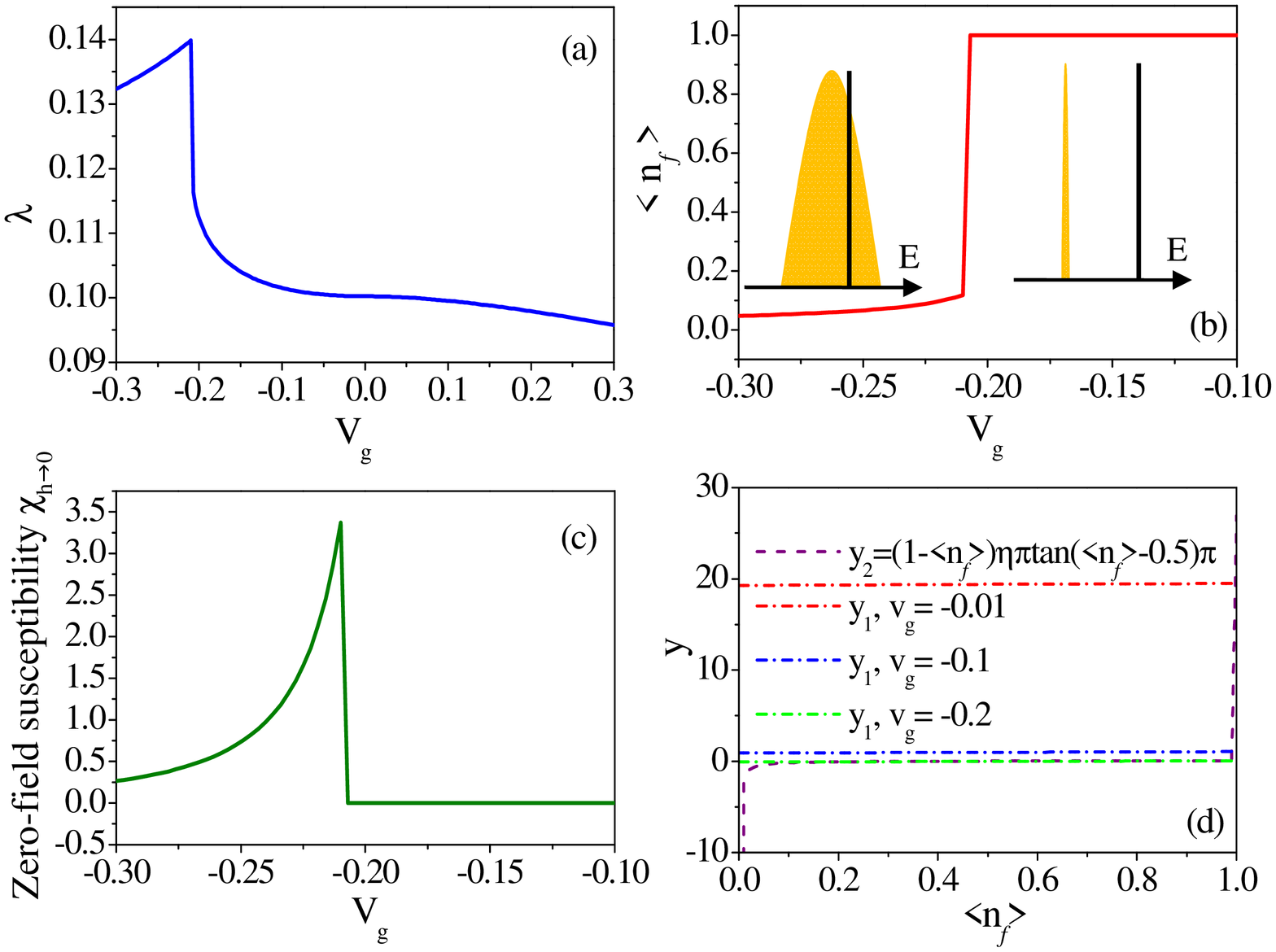}
\caption{(color online) (a) Numerically calculated $\lambda$, (b) $\langle n_{f}\rangle$, and (c) zero field susceptibility $\chi_{h\rightarrow0}$.
 (d) A graphic method to obtain  the occupation   by setting $\lambda=0.135$. The energy unit is $D/\pi$, $\varepsilon_{0}=-0.34$, $\eta=0.02$. The shaded regions of the insets in (b) indicate the broadened impurity level and the vertical lines indicate the positions of the Fermi level with a gate voltage. The left (right) inset corresponds to  lower (higher) occupation. The function in (d) is $y_{1}=\text{sgn}(v_{\text{g}})-\tilde{\varepsilon}_{0}/|v_{\text{g}}|+(1-\langle n_{f}\rangle)\eta\text{sgn}(v_{\text{g}})\ln|\pi^{2}-v_{\text{g}}^{2}|/v_{\text{g}}^{2}$.  \label{fig4}}
\end{figure}

%In Fig. (4a), the Dirac cones for up and down spins are now split by $h$, up (down) being shifted upward (downward). For up spins, the Fermi level is now lying in the %valence band (hole-like, even parity); while it is in the conduction band (electron-like, odd parity) for down-spin state. The DOS of graphene at Fermi level is however the same. The splitting is symmetric in energy manifold but asymmetric in $\Theta$ manifold. Thus nonzero $\chi$ is possible if the impurity is not fully occupied, i.e. away from the Kondo limit. In the Kondo limit, $\zeta^{2}=0$ makes the splitting is not sensible when the renormalized energy level is not close to the Fermi level. We demonstrate this by numerical selfconsistent calculations in Fig. (4b)-(4d).
Figs. (4a)-(4c) show self-consistent numerical calculations  demonstrating our above arguments.
In Fig. (4a) $\lambda$ changes  only  slightly for $v_{\text{g}}>-0.1$, meaning   the charge fluctuation is not large when the Fermi level does not reach the impurity level.  $\lambda$ increases when a sufficient negative gate voltage is applied. However, its value can not convert the bare impurity level from negative to positive (in the local moment regime in our study). In Fig. (4b), the occupation number varies with $v_{\text{g}}$. For a small negative gate voltage, the fully occupied state is still stable, a delta-function type DOS is induced (since $\zeta^{2}\approx0$) which is  schematically shown by the right insert in Fig. (4b). When the Fermi level touches  the impurity level, the charge fluctuation has a strong influence leading to a remarkable decrease in the occupation. A broadening of the impurity level occurs (linear in $v_{\text{g}}$). This strong variation in the occupation also shows up in $d$-wave superconductors \cite{zhang}; the difference to our case is that the steep decrease stems from the full occupation under insufficient gate voltage in graphene. For a comparison, in normal metals, the occupation is not complete even before the Fermi energy touches the impurity level, leading to a much  smoother change \cite{hewson}.

 It is instructive to determine the special occupations for the varying narrow  region. When $v_{\text{g}}=\tilde{\varepsilon}_{0\sigma}+\zeta^{2}\Re\Sigma_{\sigma}$, $\langle n_{f}\rangle=0.5$. When the gate voltage is lowered further, $\langle n_{f}\rangle=0.5+\frac{\Theta^{(0)}(v_{\text{g}})}{\pi}$, where $\Theta^{(0)}(v_{\text{g}})=\tan^{-1}(w(v_{\text{g}}))$ and $w(v_{\text{g}})=\frac{\text{sgn}(v_{\text{g}\sigma})}{\pi}\ln\frac{|D^{2}-v^{2}_{\text{g}\sigma}|}{v^{2}_{\text{g}\sigma}}$. In Fig. (4b), at this point, $\langle n_{f}\rangle\approx0.4$. The $\chi_{h\rightarrow0}$ is only nonzero when full occupation, i.e. $\langle n_{f}\rangle=1$, is violated which is shown in Fig. (4c) by a peak as lowering the gate voltage. To understand this strong change in the occupation, we derived the solutions of the occupation by the graphic method shown in Fig. (4d). The crossing points between the dashed and dash-dot lines delivers the solutions of the occupation. The tangent function is deformed by the relativestic linear dispersion to a  step-like function resulting in a steep change with $v_{\text{g}}$  of the occupation, which is also comprehensible from our  winding picture. Since $\zeta^{2}=0$ fixes the endpoint to $2\Theta=\pi$, the occupation can change only when the Fermi level crosses the renormalized impurity level where a sign change occurs.

\section{Discussions and interpretations}
For a deeper insight into the step-like variation of the occupation with the gate voltage, we consider the velocity of the parameters $\lambda$ and $\langle n_{f}\rangle$ with respect to $\varepsilon_{F}$ in absence of an external magnetic field.   $\lambda$ is governed by the relation
\begin{equation}
\lambda=\frac{\Im}{\pi}\sum_{\sigma}\int_{-D}^{\varepsilon_{F}}\frac{d\omega\Sigma_{0}}{\omega^{+}-\tilde{\varepsilon}_{0}-\zeta^{2}\Sigma_{0}}.
\label{lambdaint}
\end{equation}
After some algebra we find that
\begin{equation}
\frac{\partial\lambda}{\partial\varepsilon_{F}}=\frac{(\tilde{\varepsilon}_{0}-\varepsilon_{F})\mathcal{N}_{f}(\varepsilon_{F})}{\zeta^{2}},
\label{lambdaef}
\end{equation}
where $\mathcal{N}_{f}=\sum_{\sigma}\mathcal{N}_{f\sigma}$. The variation of the occupation with respect to $\varepsilon_{F}$ can be derived as
\begin{equation}
\frac{\partial\langle n_{f}\rangle}{\partial\varepsilon_{F}}=\mathcal{N}_{f}(\varepsilon_{F})=
\frac{\partial\Theta(\varepsilon_{F},\lambda,\zeta^{2})}{\partial\varepsilon_{F}}.
\label{nfef}
\end{equation}
From Eq. (\ref{nfef}), we  infer that the velocity of the occupation with a varying   gate voltage is determined by the LDOS at the Fermi level. This velocity or LDOS also describes the curvature of the manifold $\Theta$ at the Fermi level, as interpreted in the previous section. By noting the fact that the LDOS is positive, the occupation increases with raising the gate voltage above the Dirac point. When lowering the gate voltage below the Dirac point, the occupation decreases. To know how fast the velocity of  variation can be, we write  explicitly
\begin{equation}
\mathcal{N}_{f}(\omega)=-\frac{1}{\pi}\frac{\zeta^{2}\Im\Sigma_{0}}{(\omega-\tilde{\varepsilon}_{0}-\zeta^{2}\Re\Sigma_{0})^{2}+(\zeta^{2}\Im\Sigma_{0})^{2}}.
\label{ldos}
\end{equation}
When $\zeta^{2}\rightarrow0$ ($\langle n_{f}\rangle\rightarrow1$), the LDOS develops a delta function at the virtual impurity level. This is the case of half-filling. Therefore, if the virtual level is below the Fermi energy and a gate voltage is applied to drive the system away from the half-filling regime, the vanishing LDOS is manifested as a vanishing velocity of the occupation with the gate voltage, unless the Fermi level touches the virtual impurity energy. This is the reason for the behaviour observed in the  numerical calculations.
The variation of  $\lambda$ with the gate voltage, From Eqs. (\ref{lambdaef}) and (\ref{ldos}), shows a maximum at $\tilde{\varepsilon}_{0}=\varepsilon_{F}$ and  the restriction of $\zeta^{2}$ disappears. When the gate voltage is  apart from the virtual level, $\lambda$ decreases. This can also be observed  in  Fig. (4a).

\section{Summary}
In summary, we investigated the Anderson model for a magnetic adatom above one carbon atom of one monolayer of graphene. We utilized a topological method, i.e. a degree of a map and a winding number, to analyze the occupation of the impurity and the phase shift. It is found that the phase shift is accelerated by the relativistic dispersion of graphene to complete one winding or zero winding when the impurity level is respectively  below or above the Fermi energy at half-filling. %This state is even stable for non-sufficient large gate voltage demonstrated by a self-consistent numerical calculation.
The occupation  varies  dramatically from 1 (full occupation of the impurity) in a narrow range giving rise to a peak in the zero field susceptibility. The velocities of the renormalization of the impurity level and the occupation with respect to the gate voltage are worked out and
 an interpretation of the step-like variation of the occupation is provided and is consistent with the topological picture.

The work is supported by DFG and the state Saxony-Anhalt.
%\end{acknowledgement}

%


\begin{thebibliography}{}
\bibitem{footnote2}{As discussed
 in Ref. \cite{hewson}, the $\omega$-variation of the selfenergy does not contribute
 to the integral that determines $\langle n_{f}\rangle$.}
\end{thebibliography}
\end{document}